\documentclass[conference]{IEEEtran}
\IEEEoverridecommandlockouts
\usepackage{amsmath,amssymb,amsfonts}
\usepackage{algorithm}
\usepackage{algorithmic}
\usepackage{comment}
\usepackage{graphicx}
\usepackage[numbers,square,sort&compress]{natbib}
\usepackage{textcomp}
\usepackage{soul}
\usepackage{xcolor}
\usepackage{multirow}
\def\BibTeX{{\rm B\kern-.05em{\sc i\kern-.025em b}\kern-.08em
    T\kern-.1667em\lower.7ex\hbox{E}\kern-.125emX}}

\DeclareMathOperator*{\argmin}{arg\,min}    

\newcommand*\rot{\rotatebox{90}}
\providecommand{\tb}[1]{\textit{\textbf{#1}}}

\begin{document}

\title{Automated Runtime-Aware Scheduling \\for Multi-Tenant DNN Inference on GPU
\thanks{This work is partially supported by NSF CNS-2003211 and CNS-1939380.}}

\author{
	\IEEEauthorblockN{Fuxun Yu$^1$, Shawn Bray$^2$, Di Wang$^3$, Longfei Shangguan$^3$, Xulong Tang$^4$, Chenchen Liu$^2$ and Xiang Chen$^1$}
	\textit{$^1$George Mason University, Fairfax, VA, USA} \\
	\textit{$^2$University of Maryland, Baltimore County, Baltimore, MD, USA} \\
	\textit{$^3$Microsoft, Redmond, WA, USA} \\
	\textit{$^4$University of Pittsburgh, Pittsburgh, PA, USA} \\
	\textit{$^1$\{fyu2, xchen26\}@gmu.edu, $^2$\{shawnb2, ccliu\}@umbc.edu} \\
	\textit{$^3$\{wangdi, longfei.shangguan\}@microsoft.com, $^4$tax6@pitt.edu}\vspace{-3mm}
}
\maketitle

\begin{abstract}
With the fast development of deep neural networks (DNNs), many real-world applications are adopting multiple models to conduct compound tasks, such as co-running classification, detection, and segmentation models on autonomous vehicles. Such multi-tenant DNN inference cases greatly exacerbate the computational complexity and call for comprehensive collaboration for graph-level operator scheduling, runtime-level resource awareness, as well as hardware scheduler support. However, the current scheduling support for such multi-tenant inference is still relatively backward. In this work, we propose a resource-aware scheduling framework for efficient multi-tenant DNN inference on GPU, which automatically coordinates DNN computing in different execution levels. Leveraging the unified scheduling intermediate representation and the automated ML-based searching algorithm, optimal schedules could be generated to wisely adjust model concurrency and interleave DNN model operators, maintaining a continuously balanced resource utilization across the entire inference process, and eventually improving the runtime efficiency. 
    Experiments show that we could consistently achieve 1.3$\times$$\sim$1.7$\times$ speed-up, comparing to regular DNN runtime libraries (e.g., CuDNN, TVM) and particular concurrent scheduling methods (e.g., NVIDIA Multi-Stream).
\end{abstract}

\section{Introduction}
\vspace{1mm}

As deep neural networks (DNNs) have demonstrated superior performance in vast cognitive tasks~\cite{imgnet,coco,speech}, the expectations for DNN-powered intelligence have grown rapidly over the past few years.
	In addition to the real-time needs of DNN optimization regarding its deep structures and heavy workloads~\cite{runtime1,runtime2,runtime3}, recent real-world applications further require \textit{multi-tenant DNN computation} for even compound tasks~\cite{multi1, multi2, multi3}.
	For example, it is critical for an autonomous driving system to inference multiple DNN models simultaneously on the same hardware for segmentation~\cite{seg}, detection~\cite{detection}, and classification~\cite{lane}.
	And for larger-scale cases, such multi-tenant computing necessity also emerges in cloud computing clusters and industrial-level data centers for resource utilization improvement, drawing significant attention from intelligence services providers, such as Microsoft and NVIDIA~\cite{microsoft,triton,a100}.

The \tb{multi-tenant DNN inference} exacerbates the computational complexity on top of existing DNN problems.
    However, the corresponding computing support is still relatively backward.
	As the major platform for the multi-tenant inference --- current GPUs' computing strategies are still limited to traditional approaches of \textit{sequential execution} (e.g., MPI-processing~\cite{MPI}) and \textit{parallel/concurrent execution}\footnote{We treat ``parallel'' and ``concurrent'' as the same meaning in our work, according to the definition of ``concurrency'' in the NVIDIA document~\cite{stream}.} (e.g., NVIDIA multi-Stream execution~\cite{stream}).
	

\begin{figure}[!tb]
	\centering
	\includegraphics[width=3.5in]{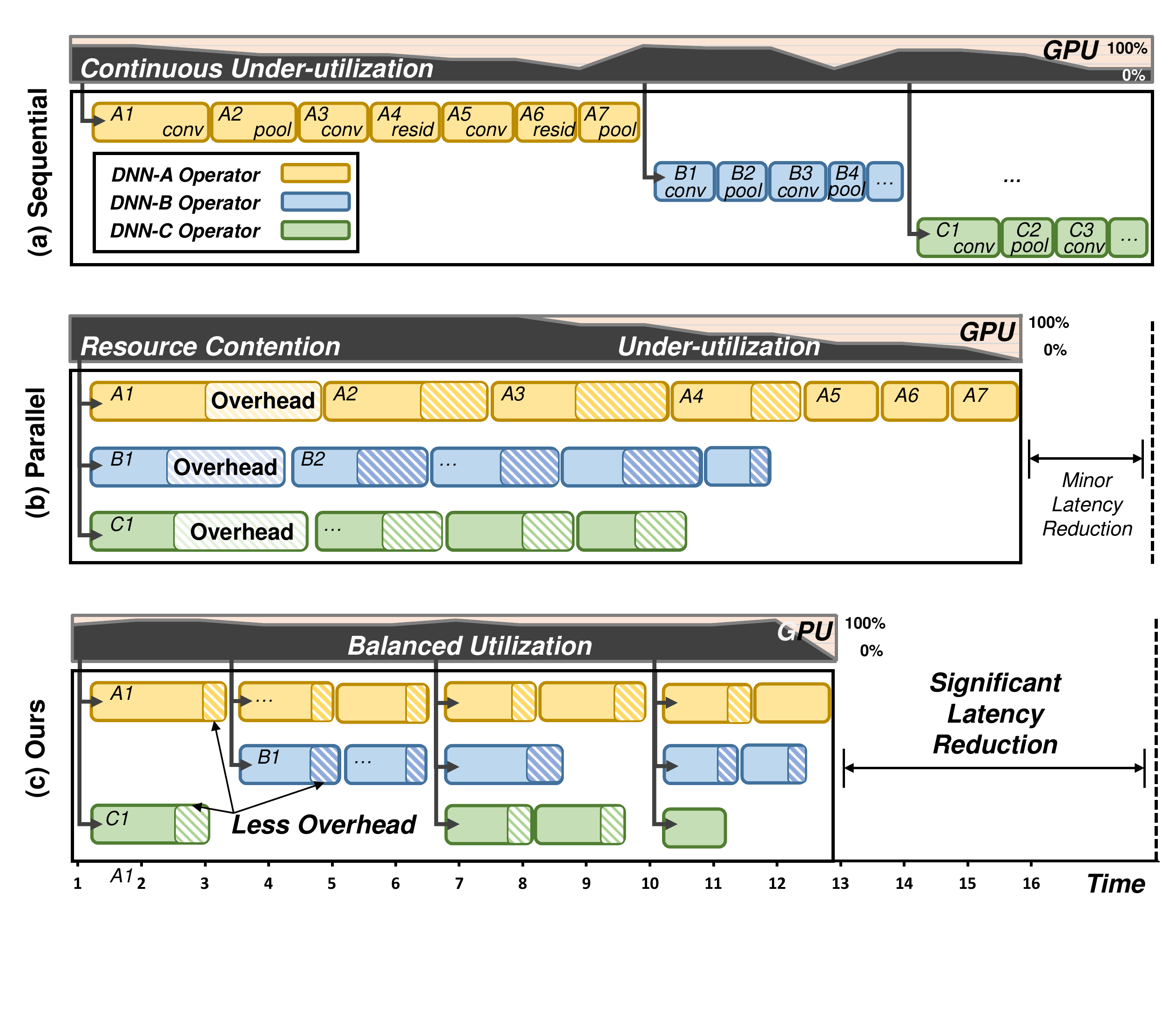}
	\vspace{-16mm}
	\caption{Scheduling for Multi-tenant DNN Inference on a Single GPU}
	\vspace{-4mm}\label{fig:1}
\end{figure}

As demonstrated in Fig.~\ref{fig:1}, these limited strategies cannot achieve satisfying performance for multi-tenant DNN inference:
	\textbf{(a)} Although the sequential execution dedicates the entire GPU's resource to each model and achieves the shortest per-model inference latency as shown in Fig.~\ref{fig:1}~(a), continuous \tb{resource under-utilization} is inevitable due to the single-operator execution (e.g., \textit{conv}, \textit{pooling}), not to mention the cumulative overall \tb{runtime latency}.
	\textbf{(b)} For the concurrent execution in Fig.~\ref{fig:1} (b), though indispensable parallelism for multiple models earns
latency optimization to a certain degree, it hasn't touched the particular computing complexity in multi-tenant DNN inference.
	Taking the first round of convolution from the three DNN models (i.e., \textit{A1}, \textit{B1}, \textit{C1}) as an example, simple parallelism would introduce considerable \tb{contention overhead} as operators can compete for computing resources simultaneously.
	While looking into later stages of this concurrent execution, GPU under-utilization strikes back due to the \tb{unbalanced scheduling} for different model depths.

\begin{figure}[!tb]
	\centering
	\vspace{-1mm}
	\-\hspace{-4mm}
	\includegraphics[width=3.5in]{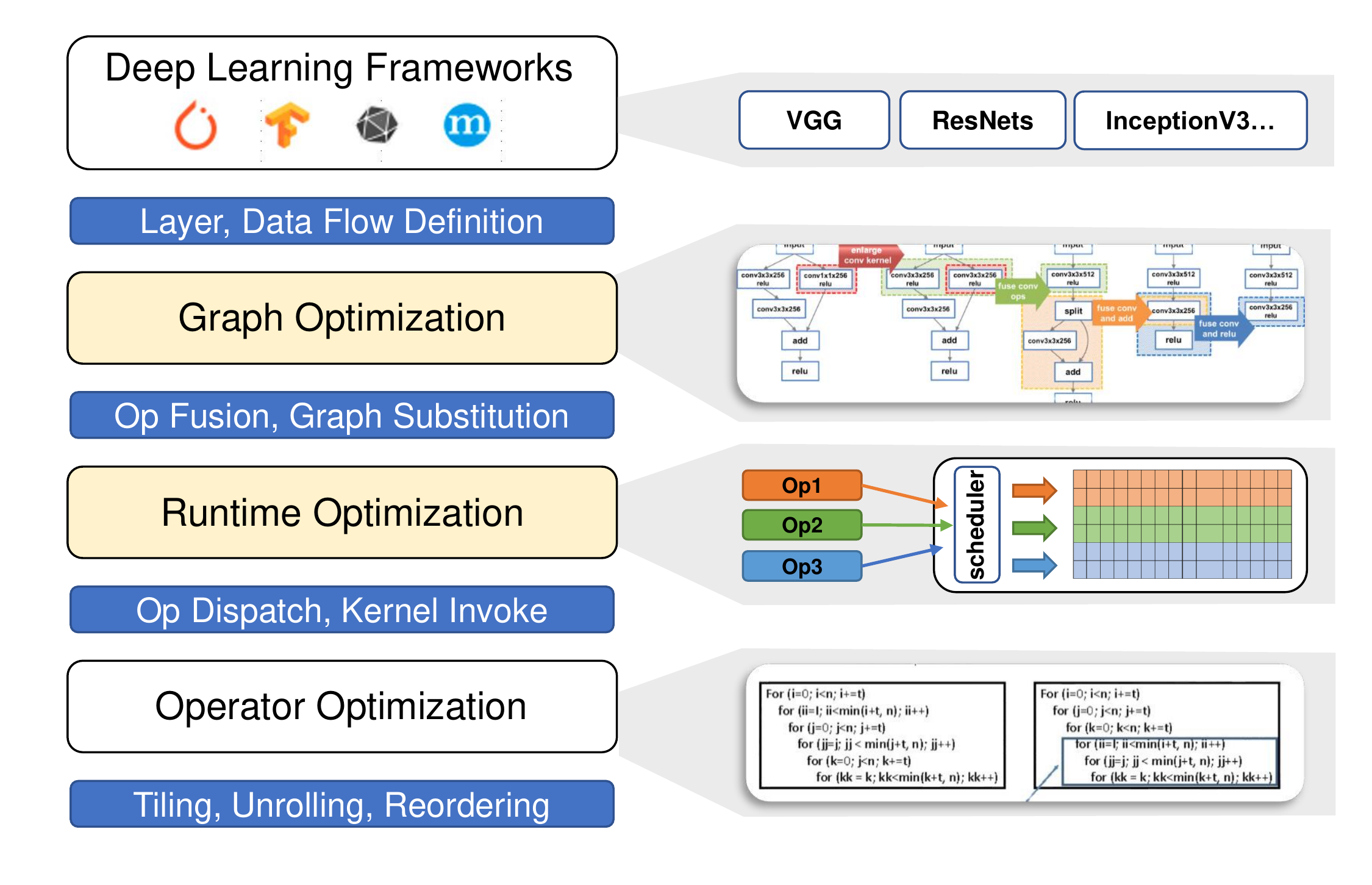}
	\vspace{-3mm}
	\caption{DNN execution stack. Our work proposes a graph- and runtime-level cross-layer scheduling framework for multi-tenant inference optimization.}
	\vspace{-3mm}\label{fig:stack}
\end{figure}

Thus, to strive for optimal runtime latency and resource utilization, the multi-tenant DNN inference raises particular GPU scheduling requirements not only for analyzing and relieving \textit{local} operator contention, but also for managing \textit{global} model concurrency balance as per model structure divergence.
	Bringing this ``\textit{local-global}'' need into the existing DNN execution stack as shown in Fig.~\ref{fig:stack}, we can see that, it calls for comprehensive collaboration from \tb{the graph-level operator scheduling}, \tb{the runtime-level resource awareness}, as well as \tb{the hardware scheduler support}.
	However, most existing DNN scheduling methods are limited in a single-level optimization scope.
	For example, many works are proposed singularly for low-level intra-operator optimization, such as loop tiling and unrolling~\cite{tvm,xla,tensorrt}; Similarly, many graph-based scheduling works focus only on high-level inter-operator fusion/substitution optimization~\cite{taso,mlsys,transferable}.
	As a result, neglecting one or the other, these methods fail to meet the cross-level scheduling requirement by the multi-tenant DNN inference.


In this work, we propose a \textit{runtime-aware scheduling framework for efficient multi-tenant DNN inference on GPU}, which \textit{automatically} coordinates concurrent DNN computing in different execution levels.
	As shown in Fig.~\ref{fig:1}~(c), the proposed method could take both the local operator contention and the global model structural divergence into consideration.
	The final scheduling method wisely adjusts model concurrency by interleaving operators for less contention overhead, maintaining a continuously balanced resource utilization across the entire inference process, and eventually improving the runtime efficiency.
	To achieve such a scheduling target, we make the following contributions:

\begin{itemize}

	\item	We first \textit{abstract the multi-tenant DNN inference scheduling} as a fine-grained concurrency control problem.
	Incorporating the GPU multi-stream and synchronization mechanisms, multiple concurrency control levels are identified in the GPU inference flow to provide the fundamental support for the scheduling optimization;

	\vspace{1mm}
	\item Based on the problem abstraction, \textit{a unified scheduling Intermediate Representation (IR)} is specified to formulate the scheduling factors by taking both graph-level and the runtime-level execution mechanism into consideration, and eventually build a structural search space for the final scheduling optimization;

	\vspace{1mm}
	\item In the established scheduling search space, we transform multi-tenant scheduling into an optimization problem and propose an automated ML-based searching algorithm to find the optimal scheduling strategy on GPU.
	Specifically, the GPU runtime resource is profiled and adopted as the searching cost, granting the whole solution with expected runtime awareness.

\end{itemize}

We conduct extensive experiments across a wide range of multi-tenant inference scenarios.
	The results show that our method could consistently achieve 1.3$\times$$\sim$1.7$\times$ acceleration than the common deep learning runtime libraries (e.g., CuDNN, TVM) and other concurrent scheduling methods (e.g., NVIDIA Multi-Stream).
	Meanwhile, benefited from the end-to-end search method design, our method could be easily applied onto 10s of multi-tenant combinations and GPU platforms with short search time ($\sim$2mins), demonstrating the great scalability of our automated scheduling framework.



\vspace{2mm}
\section{Backgrounds and Motivation}
\vspace{2mm}

\subsection{Cross-Level Scheduling through DNN Execution Stack}

We first expand the backgrounds of DNN execution stack as shown in Fig.~\ref{fig:stack}, that is composed of multiple architecture levels~\cite{tvm}:
	\textbf{(a)} The top \textit{framework level} includes different deep learning development frameworks, such as TensorFlow and PyTorch, that define various DNN model structures.
	\textbf{(b)} The \textit{graph level} untangles model structures to abstract individual operators and identify the data processing flow as directed acyclic graphs (DAG)~\cite{taso}.
	Graph-based optimization is thus introduced into this level to achieve operator fusion and sub-graph substitution, and therefore reduce {memory access/operator invoking} overheads, etc.
	\textbf{(c)} Down to the \textit{runtime level}, it controls when and how operators are dispatched onto physical computing units and is critical in our balanced resource utilization.
	This is usually done by the native black-box GPU scheduler, but we could leverage certain APIs to adjust the dispatching results.
	In our work, we use the ``stream''~\cite{stream} and ``synchronization''~\cite{sync} APIs to achieve fine-grained {operator} concurrency control as we will show later.
	\textbf{(d)} The \textit{operator level} is the bottom level that conduct per-operator execution, such as tiling, unrolling, reordering, etc., to improve the computing efficiency.
	Such {intra-operator} optimization has a distinct scope and is orthogonal to the concurrent operator scheduling, and thus is not considered in this work.

\begin{figure*}[!t]
	\centering
	\vspace{-2mm}
	\includegraphics[width=7in]{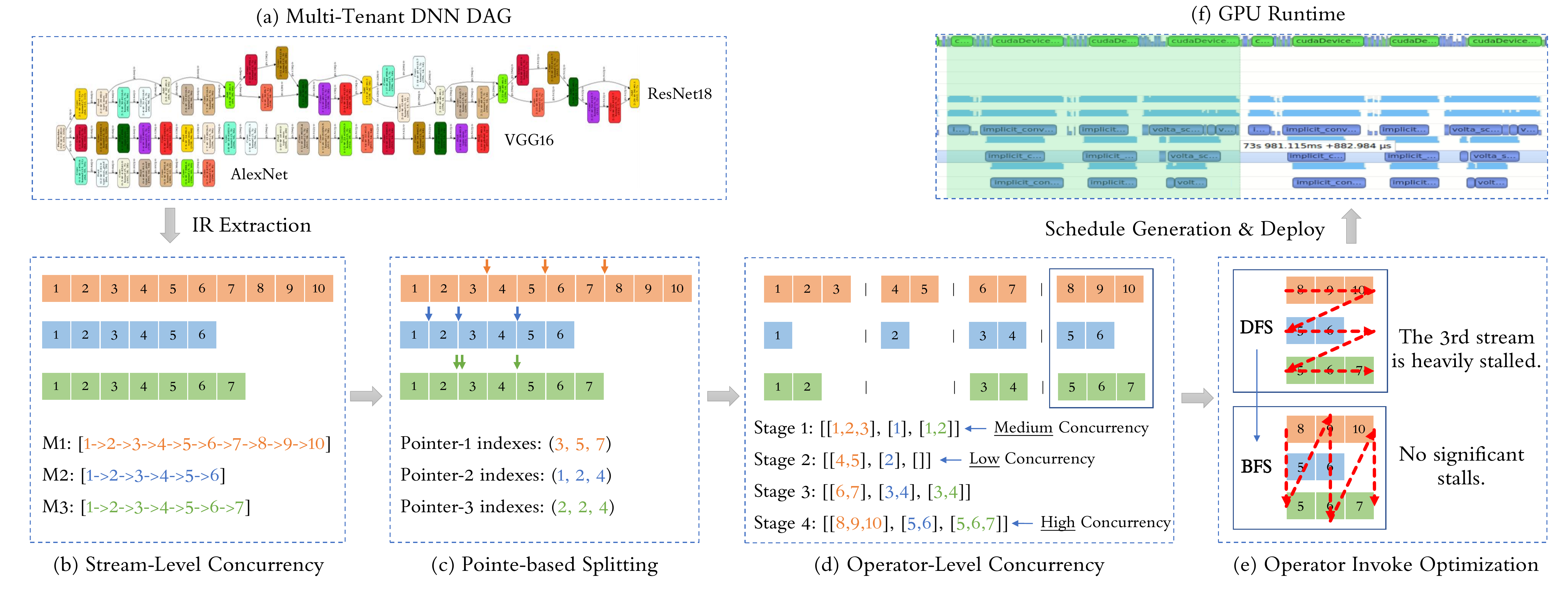}
	\vspace{-2mm}
	\caption{Overview of Our Proposed Automated Scheduling Strategy Search Framework.}
	\vspace{-4mm}
	\label{fig:framework}
\end{figure*}

\vspace{0.5mm}
\tb{Motivation:}
	As existing single-level works (e.g., graph-alone, operator-alone) can hardly offer comprehensive solution, we aim to bridge different levels in this work and build a cross-level scheduling framework to improve the multi-tenant inference performance from graph to runtime.





\subsection{Resource Contention in Multi-Tenant DNN Inference}
	Resource contention is a specific challenge emerging in the multi-tenant inference.
	Different from single-model inference which usually faces under-utilization issues, resource contention reflects the operator competition for limited hardware resources.
	Specifically, there are two types of contention that are commonly known: \textit{computing contention} and \textit{memory contention}.
	Generally, convolutional operators tend to be compute-bound as it involves mostly FLOPs-intensive computing, while the operators like pooling/residual connections are usually memory-bound.
	When executed concurrently, same type of operators can saturate the corresponding resources, and cause slower execution speed due to the limited size of shared memory and L1 cache, memory bandwidth, etc.

\vspace{0.5mm}
\tb{Motivation:}
	Thus, multi-tenant scheduling is a finer concurrency control problem, which not only concerns under-utilization but also contention issues.

\subsection{Scheduling Complexity in Multi-Tenant DNN Inference}
	Moreover, multi-tenant inference optimization have a much larger complexity:
	\textbf{(a)} From a spatial perspective for inference parallelism, on each single execution stage, the overall amount of operators from different models is much more significant than a single-model scheduling scope.
	\textbf{(b)} From a temporal perspective, through the whole execution process, different model structures and depths also raise considerable scheduling challenges to maintain consistent and balanced resource utilization and improve the runtime latency.
	Therefore the design/search space of the scheduling strategy for multi-tenant DNN inference becomes ever complex, and conventional manual tuning or heuristic-based methods are hard to scale and reach satisfying performance. 

\vspace{0.5mm}
\tb{Motivation:}
	In addition to other design motivations, we will eventually solve this problem by proposing an efficient search space representation and leverage automated ML-based methodologies to coordinate massive operators for optimal resource utilization and runtime latency.

\vspace{1mm}
\section{The Scheduling Framework}
\vspace{1mm}

\subsection{Fine-grained Scheduling Problem Abstraction}

This work targets at efficient multi-tenant DNN inference on GPUs. 
%
Considering the applications such as autonomous driving systems, we specify it as a compound task consisting of $N$ independent DNN models sharing the same input for different inference sub-tasks.
	Demonstrated as Fig.~\ref{fig:framework}~(a), each DNN inference sub-task consists of a series of operators, such as \textit{conv}, \textit{bn}, \textit{relu}, \textit{pooling}, etc, which must be performed in certain order according to the data flow dependency.
	While across DNN models, operators are independent and thus could be flexibly scheduled with certain degrees of concurrency.
Our optimization objective is to minimize the overall latency of $N$ inference sub-tasks, which is the overall time from the earliest starting time of the tasks to the latest ending time.


The key of multi-tenant scheduling is to manage the concurrency for consistent and balanced resource utilization.
    Therefore, we abstract the multi-tenant DNN inference scheduling as a fine-grained  concurrency control problem through the following steps:
\textbf{(a)} \textit{Achieving the \tb{stream-level} concurrency}:
	We allocate one \textit{GPU processing stream} for each model to achieve the concurrency (Fig.~\ref{fig:framework}~(b)).
	However, even with certain concurrency, native GPU stream-based scheduling dispatches operators without dedicated scheduling management.
\textbf{(b)} \textit{Finer-grained} \tb{stage splitting}:
	To achieve finer-grained operator-level concurrency control, we insert synchronization barriers, namely \tb{pointers}, to split each stream's operator sequence into several shorter stages (Fig.~\ref{fig:framework}~(c)).
	Such stage splittings ensure the operators to only share the assigned resources in the same stage, thus supporting the stage-level concurrency control.
\textbf{(c)} \tb{Stage-level} \textit{concurrency control}:
	By adjusting where the pointers are inserted, we could control how many operators are assigned in each stage.
	This enable us to reduce or increase the concurrency in a fine-grained manner to manage the resource utilization (Fig.~\ref{fig:framework}~(d)). 
\textbf{(d)} \tb{Intra-stage} \textit{operator invoking optimization}:
    After deciding the scheduling strategy, our final step is the scheduling deployment. During this implementation, we also optimize the operator invoking logic to prevent the invoking overhead of early streams from stalling later ones (\ref{fig:framework}-e), as we will introduce later.

\subsection{Unified Intermediate Representation Design}
\label{sec:ir}

As a multi-tenant DNN inference task consists of $N$ parallelable models: $M_1, M_2, ..., M_N$. 
We represent each DNN model by one stand-alone operator sequence\footnote{For multi-branch models like ResNets, we also serialize the operators into one sequential sequence as their \textit{intra-model concurrency} is limited. Such a representation enables us to better optimize the \textit{inter-model concurrency} in the multi-tenant inference scenario.}: 
\begin{equation}
\small
\begin{split}
\text{M}_1: ~&[1, ~2, ~..., ~a],\\
\text{M}_2: ~&[1, ~2, ~..., ~b],\\
\text{M}_N: ~&[1, ~2, ~..., ~c],\\
\end{split}
\label{eq:sequence}
\end{equation}
\normalsize
where $M$ indicates a DNN model, each number in one list indicates one operator's index, and $a$ (or $b,c$) is the largest index of the DNN's operators. 

\noindent \tb{{Stream}}: To satisfy the sequential dependency per model, we assign each model to one stand-alone \textit{{GPU processing stream}}:
\begin{equation}
\small
S_i \leftarrow ~\text{M}_i, ~~~i \in (1, 2, ..., N),
\end{equation}
\normalsize
where $S_i$ indicates the $i$-th stream. 
An example with three streams is shown in Fig.~\ref{fig:framework} (b).
Operators in one stream can only be launched sequentially, while operators in different streams could be executed concurrently. 

The multi-stream mechanism enables the maximum concurrency of DNN inference streams. 
	However, as aforementioned, scheduling by streams alone can only have \textit{stream-level concurrency control}, which is still coarse-grained and does not suffice to manage each operator's associated concurrency during its life span. 
	To control the concurrency in a finer granularity, we then use synchronization barriers to split each stream's full sequence into several shorter stages.

\vspace{2mm}
\noindent \tb{{Pointer}}: We use \textit{pointers} to annotate the appropriate positions where we insert synchronization barriers.
	An illustrated pointer-based stage splitting is shown in Fig.~\ref{fig:framework} (c)(d).
	Taking the first stream as an example,
	a pointer set with three pointers divides the first stream sequence into four shorter ones:
\begin{equation}
\begin{split}
\rho_1: (\textbf{3, 5, 7}) ~+~
\text{S}_1: [1,2,3,...,9,10] = \\
\text{S}_1^{'}: [1,2,3], [4,5], [6,7], [8,9,10],
\end{split}
\label{eq:pointer}
\end{equation}
where $\rho_1$ is the pointer indexes, $S_1$ is the original operator sequence,  and $S_1^{'}$ is the split sequence with synchronization barriers inserted.
	Each pointer set splits one stream sequence into several shorter ones, thus enabling a finer-grained concurrency scheduling.

\vspace{2mm}
\noindent \tb{{Stage}}: Between each two pointers, the launched operators form a \textit{stage}.
	Due to the sync barriers, all operators in the same stage must all finish so as to step into the next stage.
	Thus, by controlling how many operators are launched in each stage, we could precisely manage the concurrency in the most fine-grained operator level.
An example is given in Fig.~\ref{fig:framework} (d).
By inserting the first synchronization barrier, we could enable six operators to concurrently execute in the first stage:
\begin{equation}
\small
\text{Stage 1}: [S_1{(1,2,3)}, ~~S_2{(1)}, ~~S_3{(1,2)}].
\end{equation}
\normalsize
By contrast, we could also reduce the concurrency in the second stage by assigning no operators in the third stream:
\begin{equation}
\small
\text{Stage 2}: [S_1{(4,5)}, ~~S_2{(2)}, ~~S_3{(\text{None})}].
\end{equation}
\normalsize
Similarly, all stages can be generated with a desired concurrency, thus enabling \textit{operator-level concurrency control}.

\vspace{2mm}
\noindent \tb{{Schedule}}: The final scheduling strategy is composed of multiple stages in the synchronization barriers' ordering, which is represented as a multi-stage nested list:
\begin{equation}
\text{Schedule} ~\tau: [\text{Stage 1}, ~\text{Stage 2}, ~\text{Stage 3}, ...],
\label{eq:schedule}
\end{equation}
where $\tau$ indicate the composed scheduling strategy, which can have multiple stages, depending on the number of synchronization barriers (i.e., pointers) we used to split each stream sequence.
Fig.~\ref{fig:framework} (d) shows an example that uses three sync pointers for four stages.
More synchronization enables finer-grained concurrency control, but at the price of potentially higher synchronization overhead.

\begin{figure}[!b]
	\centering
	\vspace{-5mm}
	\includegraphics[width=3.4in]{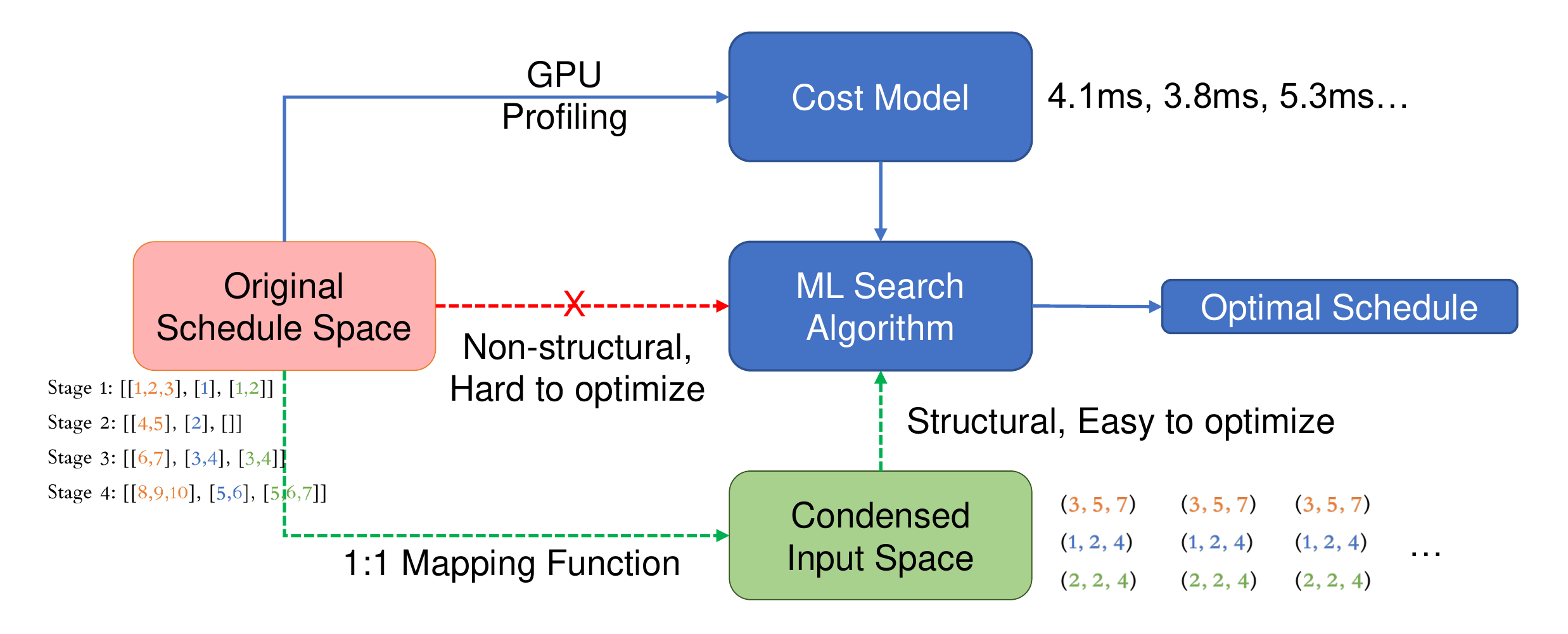}
	\vspace{-5mm}
	\caption{The automated scheduling search framework overview.}\label{fig:ml}
\end{figure}

\subsection{Automated Scheduling Search}

The IR design explicitly defines the scheduling factor and the corresponding strategy for multi-tenant GPU inference.
    However, it is still challenging to identify the particular scheduling controls given various compound tasks with uncertain DNN structures. 
    As aforementioned, considering the complexity, manual schedule tuning can take considerable efforts and also cannot scale with more complicated models' combination and varied GPU platforms.
	Therefore, we propose to use an ML-based search approach to solve the scheduling problem in an automated manner.


\vspace{2mm}
\noindent \tb{Formulation:} 
Formally, our primary search target is to find an optimal scheduling strategy that yields the lowest latency:
\begin{equation}
\tau^{*} = \argmin_{\tau}~ {f}(\tau), ~~~\text{for}~ \tau \in D_{\tau},
\label{eq:org_search}
\end{equation}
where $\tau^{*}$ is the optimal scheduling strategy, $f$ is the cost model that evaluates the latency of the current schedule $\tau$, and $D_{\tau}$ is the search space of all potential schedules.
    Specifically, to solve this search problem, three basic components need to be clarified, namely, the search space, the cost model and the searching algorithm.

\vspace{2mm}
\noindent \tb{Search Space} is supposed to enumerate all possible scheduling strategy candidates.
    To represent such a search space, we adopt the scheduling factors from the proposed IR design (i.e., streams, pointers and stages).

As defined in Eq.~\ref{eq:schedule}, $\tau$ can be formulated as a nested list and can be treated as a graph-level scheduling strategy.
	Although such a nested list is easy to understand and facilitates the deployment process onto GPU, the list-based search space $D_{\tau}$ is non-structural with varied list lengths and can be hard to directly optimize.
	To solve this problem, we leverage the one-to-one mapping property between pointer indexes $\rho$ and the schedule lists $\tau$, and shrink the search space to a lower-dimensional pointer index matrix by building an 1:1 schedule mapping function, as shown in Fig.~\ref{fig:ml}:
\begin{equation}
\begin{split}
& \rho^{*} = \argmin_{\rho}~ {f}(\tau), \\ 
& \text{s.t.} ~~\tau = T(G, \rho), ~~\text{for}~ \rho \in D_{\rho}.
\label{opt_search}
\end{split}
\end{equation}
Here the scheduling generation function $T(\cdot)$ generates one schedule $\tau$ based on two inputs: the graph $G$ and the pointer matrix $\rho$. 
	As $G$ is usually fixed in a given task, the schedule generation function $T(\cdot)$ maps each pointer matrix to one schedule.
	Thus, searching schedule could be transformed to searching the pointer index matrix, the latter of which has a much more structural input space.
	By such transformation, we could thus greatly reduce the optimization difficulty.

\begin{algorithm}[!b]
   \caption{Coordinate Descent Search Algorithm.}
   \label{alg:1}
\begin{algorithmic}[1]
   \STATE {\bfseries Input:} The IR of $N$ models $M[N]$, the number of pointers in each model $P$, the rounds of search $R$.
   \STATE {\bfseries Output:} The optimal pointer matrix $\rho~[N, P]$.
   \STATE Initialize a dictionary $D$\{schedule:cost\} to store records.
   \FOR{rounds $r=1$ {\bfseries to} $R$}
	\FOR{model $i=1$ {\bfseries to} $N$}
	   \STATE Sample $M$ candidates $\rho_{1:M}[i]$ for the $i$-th row.
	   \FOR{the $m$-th candidate $\rho_{m}[i]$}
	   \STATE Profile the latency $lat_{m}$ by multiple runs.
	   \STATE Append \{$\rho_{m}:lat_{m}$\} to the records $D$.
	   \ENDFOR
	   \STATE Update the $i$-th row $\rho[i]$ of pointer matrix to the one with the lowest latency by $\argmin(lat_{m})$.
	 \ENDFOR
	 \ENDFOR
	  \STATE Sort the global records $D$ by the profiled latency. 
	  \STATE \textbf{Return} the schedule $\rho$ with the globally lowest latency.
	  
\end{algorithmic}
\end{algorithm}

\vspace{2mm}
\noindent \tb{Cost Model}:
With the search space defined, we then require a cost model $f(\tau)$ to evaluate the performance of each schedule candidate.
	There are two major ways to construct the cost model: modeling-based or profile-based.
	The modeling-based method~\cite{mlsys} builds hardware-specific modeling to estimate the real runtime performance, which is efficient but can be inaccurate in complex scenarios.
	The profiling-based method~\cite{tvm} is more accurate but requires physical hardware execution, which can be more time-consuming if the search space is very large.

In this work, we use the profiling-based cost model since our empirical case study shows that, our searching time can be maintained at small scale ($\sim$mins) benefited from our dedicated search space abstraction.
	Therefore, the profiling-based model can give accurate runtime-aware performance cost and lead to better search performance in our case.
	For the cost model implementation, we leverage our built infrastructure, which could efficiently generate and deploy each candidate schedule onto the target GPU and obtained the profiled latency during multiple averaged runs.
	The averaged latency is then used as the cost of each candidate schedule.

\vspace{2mm}
\noindent \tb{Search Algorithm}:
With the input space and cost model defined, we could then use ML-based methods to search for the optimal schedule with the minimal latency.

\begin{figure}[!b]
	\centering
	\vspace{-4mm}
	\includegraphics[width=3.5in]{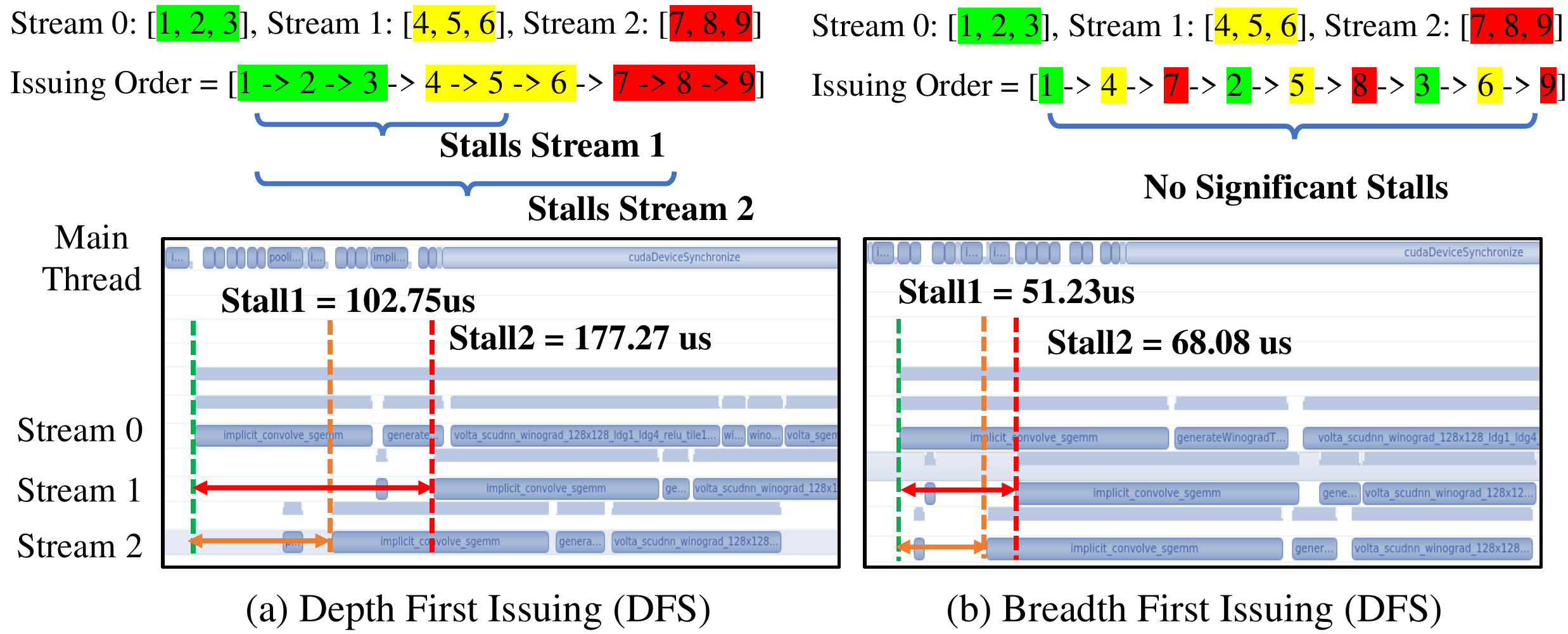}
	\caption{Long-sequence operator invoking by DFS can significantly stall other streams. We optimize the scheduler logic to BFS issuing to reduce such stall.}\label{fig:bfs}
\end{figure}

In this work, we mainly implement two search algorithms, the random search and the coordinate descent search.
The random search method samples scheduling solutions (different pointer matrices) randomly from the search space and profiles their latency as the cost.
	A memory module will record all schedules and costs, and after certain rounds of search, the algorithm will return the schedule with the lowest latency.
	As we will show later, though the random search algorithm is simple, it could greatly reduce the multi-tenant runtime latency by large margins, highlighting the advantages of our problem abstraction and the search framework design.

Based on a similar process, the coordinate descent search algorithm improves the sampling efficiency by adopting a coordinate-alternated search philosophy.
	The overview of the coordinate descent search algorithm is shown in Algorithm~\ref{alg:1}.
	It treats different streams' pointer index vectors (rows in the pointer matrix) as different coordinates.
	Then it alternatively finds the optimal pointer index vector for each coordinate, during when other coordinates' solution are kept as the previous optimal one.
	The optimal pointer index vectors for all streams are updated for each round, and after certain rounds, the algorithm returns the optimal schedule from all previously searched schedules.
	Generally, the coordinate descent search algorithm could yield slightly performance than random search algorithm.
	But both methods could yield near optimal schedule solutions within short time, as we will evaluate later.
	
\begin{figure*}[!tb]
	\centering
	\includegraphics[width=6.5in]{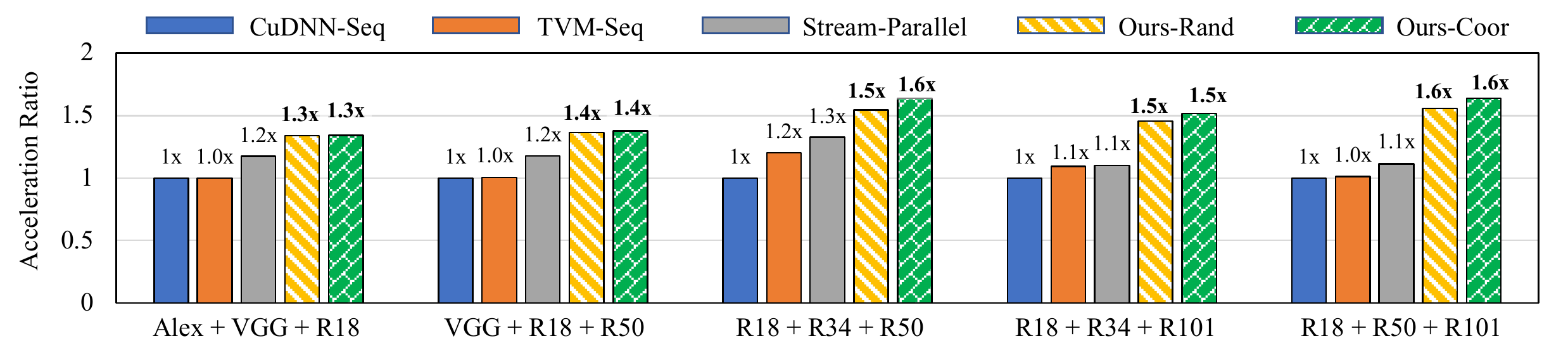}
	\vspace{-1mm}
	\caption{Runtime Performance of the Proposed Automated Scheduling Framework. We mainly compare the acceleration ratio with three baselines: CuDNN-based sequential execution (CuDNN-Seq), TVM-based sequential execution (TVM-Seq), Stream-based parallel execution (Stream-parallel). Here Ours-R and Ours-G denotes the performance of our framework with random search and coordinate descent search. Test Platform: Titan V GPU.}
	\vspace{-5mm}
	\label{fig:exp1}
\end{figure*}

\subsection{Implementation Optimization}
After determining the optimal schedule, we can deploy the schedule onto the GPUs.
    This is done by invoking the GPU kernels according to each stage's IR.
	In multi-stream execution, the operator invoking is controlled by a main thread and invoking each operator takes a small duration of time.
	Although individually small, an inappropriate invoking sequence can also influence the latency, especially in high-concurrency stages with many operators.

Fig.~\ref{fig:bfs} (a) showcase one example of operator invoking-caused stall.
	The default scheduler utilizes a depth-first (DFS) issue logic that issues all operator sequentially in one stream to ensure the operator dependency is maintained, and then iterates overall all streams.
	However, when there are multiple operators in the beginning streams, operators in the later streams can be significantly stalled due to the accumulated operator invoking overhead.
To relieve such stall, we optimize the default DFS logic into a breadth-first (BFS) strategy.
	Fig.~\ref{fig:bfs} (b) illustrates the BFS logic, which issues one operator from each stream interleavingly, and then iterates until all operators are issued.
	In such cases, all streams get similar invoking priority, and the operator dependency is also maintained in each stream.
	As a result, we could greatly reduce the operator invoking overhead for the later streams.
	Fig.~\ref{fig:bfs} shows an example which we could reduce the stall from $102.75\rightarrow51.23$us and $177.27\rightarrow51.23$us for \textit{Stream-1} and \textit{Stream-2}.
\vspace{1mm}
\section{Experimental Evaluation}


\begin{table*}[!tb]
\centering
\caption{\textbf{Scalability Evaluation (BS=1, 224X224, GPU: Titan-V w/ Volta Arch, Latency: ms})}
\renewcommand\arraystretch{1.5}
\setlength{\tabcolsep}{3.6mm}{
\begin{tabular}{ccccccc}
\hline
\multicolumn{1}{c}{} \#Models         & Names     & CuDNN-Seq & TVM-Seq & Stream-Parallel & Ours-R & Ours-C \\ \hline
\multirow{4}{*}{\rot{2$\times$models}}   & VGG + R18     & 3.989     & 3.898   & 3.638      & 3.096 \textbf{(1.29$\times$)}  & 2.912 \textbf{(1.37$\times$)}  \\ \cline{2-7} 
                              & R18 - R34     & 4.673     & 3.453   & 3.743      & 3.382 \textbf{(1.38$\times$)}  & 3.128 \textbf{(1.49$\times$)}  \\ \cline{2-7} 
                              & R34 + R50     & 6.688     & 5.785   & 5.449      & 4.725 \textbf{(1.41$\times$)}  & 4.478 \textbf{(1.49$\times$)}  \\ \cline{2-7} 
                              & R50 + R101    & 10.75     & 10.435  & 8.588      & 8.385 \textbf{(1.28$\times$)}  & 8.203 \textbf{(1.31$\times$)}  \\ \hline
\multirow{2}{*}{\rot{3$\times$}} & VGG + R18 + R50 & 7.674     & 7.637   & 6.522      & 5.639 \textbf{(1.36$\times$)}  & 5.587 \textbf{(1.37$\times$)}  \\ \cline{2-7} 
                              & R18 + R34 +R50 & 8.344     & 6.949   & 6.301      & 5.404 \textbf{(1.54$\times$)}  & 5.096 \textbf{(1.63$\times$)}  \\ \hline
{\rot{5$\times$}}              & VGG + R18 + R34 +R50 + R101    & 17.962    & 16.742  & 12.848     & 10.91 \textbf{(1.65$\times$)}  & 10.42 \textbf{(1.72$\times$)}  \\ \hline
\end{tabular}}
\label{table:scale}
\vspace{-4mm}
\end{table*}

\subsection{Experiment Setup}

\vspace{1mm}
\noindent \tb{Model Zoo for Multi-Tenant Combination:} We construct various multi-tenant scenarios by leveraging the following neural network models: AlexNet ($Alex$), VGG16 ($VGG$), ResNet18 ($R18$), ResNet34 ($R34$), ResNet50 ($R50$) and ResNet101 ($R101$).
	These models have distinctive model depths with operator numbers varying from $7\sim20$ to $86\sim216$.
	In addition, operators from different models also have particular computing and memory requirements.
	For example, the convolution operators have a wide range of computing complexity, e.g., from $32\sim64$ filters per layer to $256\sim512$ filters per layer.
	Therefore, each different multi-tenant combination based on the above models will pose its unique resource utilization imbalance challenges and has distinctive optimal scheduling strategies, mimicking the varied and complex multi-tenant scenarios of real-world applications.

\vspace{2mm}
\noindent \tb{Evaluation and Comparison Baselines:} Three popular baseline scheduling strategies are considered.
\begin{itemize}

	\item CuDNN-Seq~\cite{CuDNN}: The default strategy supported by the NVIDIA CuDNN library, which runs the multi-tenant inference sequentially;

	\vspace{1mm}
	\item TVM-Seq~\cite{tvm}: A operator-level optimization method that adopts the TVM library~\cite{tvm} to search for the optimal kernel for each operator. However, without runtime support, it can only run these kernels sequentially;

	\vspace{1mm}
	\item Stream-Parallel~\cite{stream}: The concurrent execution strategy from native GPU multi-stream support~\cite{stream}. It assigns models to different streams and leverages the default GPU scheduler to schedule the execution sequence.

\end{itemize}

\vspace{2mm}
\noindent \tb{Inference Setup:} We conduct neural network inference on ImageNet~\cite{imgnet} that has an image scale of 224x224x3 with single batch size to mimic the inference in practical applications such as autonomous driving.
	Two NVIDIA GPU platforms are utilized: Titan V of Volta architecture, P6000 of Pascal architecture. 
	For all latency measurement, we record the averaged latency (ms) by profiling the same number of runs for our method and the baselines.

\subsection{Speed-Up Evaluation}
We first compare the inference latency of the baselines and our methods.
The results are shown in Fig.~\ref{fig:exp1}.
All methods' latency is normalized by the CuDNN-Seq baseline to show the relative acceleration ratio.
Five multi-tenant settings, which cover a wide range of multi-tenant combinations are built up.
For example, $Alex+VGG+R18$ which is a relatively simple ones (10$\sim$30 operators), and $R18+R50+R101$ whose operator numbers can over 200 is the most complex one, \emph{etc}. 
	For our method, we show both search algorithms' performance in our framework -- the random search (\textit{Ours-Random}) and coordinate descent search (\textit{Ours-Coor}).


\vspace{2mm}
\noindent \tb{Overall Speed-up}: It can be observed that our scheduling framework could consistently yield $1.3\times\sim1.6\times$ speed-up compared to the sequential baselines across all five model combinations.
	Although the Stream-Parallel solution also yields a certain speed-up than CuDNN-Seq, its acceleration ratio is only $1.1\times\sim1.3\times$, which is much less than ours.

\vspace{2mm}
\noindent \tb{Higher Speed-up in Highly Non-balanced Scenarios}: 
	It is worth noting that our method achieves the highest acceleration ratio, i.e., \textbf{$1.5\times$} and $1.6\times$, on the two most challenging scenarios $R18 + R34 + R101$ and $R18 + R50 + R101$.
	However, the Stream-Parallel performs poorly (only $1.1\times$) in these two settings.
The reason is that such two multi-tenant combinations introduce extremely distinctive model lengths from 29 operators (ResNet18) to 200 operators (ResNet101), which brings significant resource imbalance between early and later stages across the entire processing.
	The native hardware scheduler in Stream-Parallel cannot take this into consideration and push all operators into the beginning stages, and thus cannot balance the resource utilization effectively.
	Therefore, it can only reach limited acceleration ratio.

\begin{figure}[!b]
	\centering
	\vspace{-5mm}
	\includegraphics[width=3.4in]{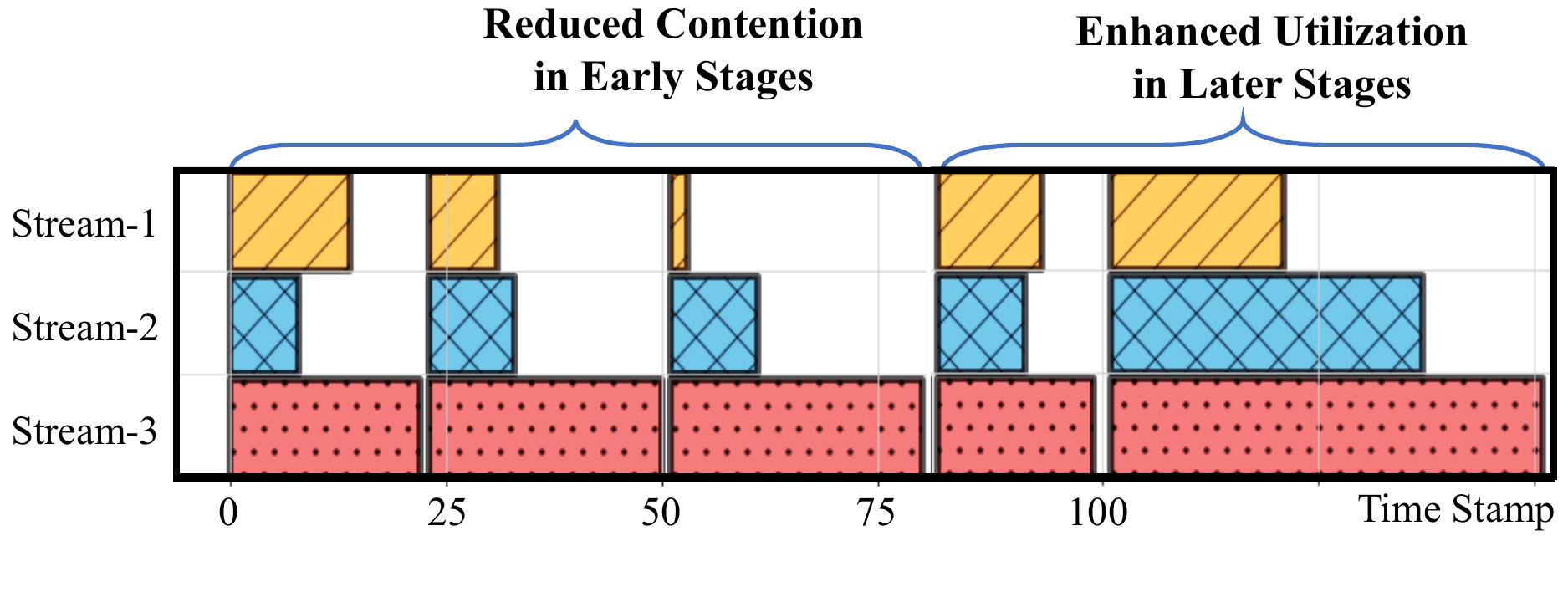}
	\vspace{-6mm}
	\caption{Illustration of our Resource Balance Mechanism: Our method could find a balanced schedule to avoid both contention and under-utilization, thus achieving better performance than sequential and native parallel solutions.}\label{fig:gatt}
\end{figure}

\vspace{1mm}
\noindent \tb{Illustration of our Resource Balance Mechanism}: 
In contrast to the hardware scheduler in Stream-Parallel, our method could effectively find a better scheduling solution via our pointer-based barrier insertion and automated search algorithm and hence achieves higher speed-up in the highly unbalanced scenarios.
We visualize the kernel invoke timeline of our scheduling strategy on the $R18 + R50 + R101$ scenario, as is depicted in Fig.~\ref{fig:gatt}, to reveal the mechanism. 
	The number of operators issued in each stage is symbolically denoted by the length of each colored block.
    The results show that our searched scheduling could effectively reduce the number of operators in the early stages to avoid potential resource contention and leave more operators into the later stages to enhance resource utilization.
	As such, our scheduling enables optimal resource utilization and finally achieves significantly lower latency performance than the Stream-Parallel solution.

\begin{figure}[!b]
	\centering
	\vspace{-5mm}
	\includegraphics[width=3.4in]{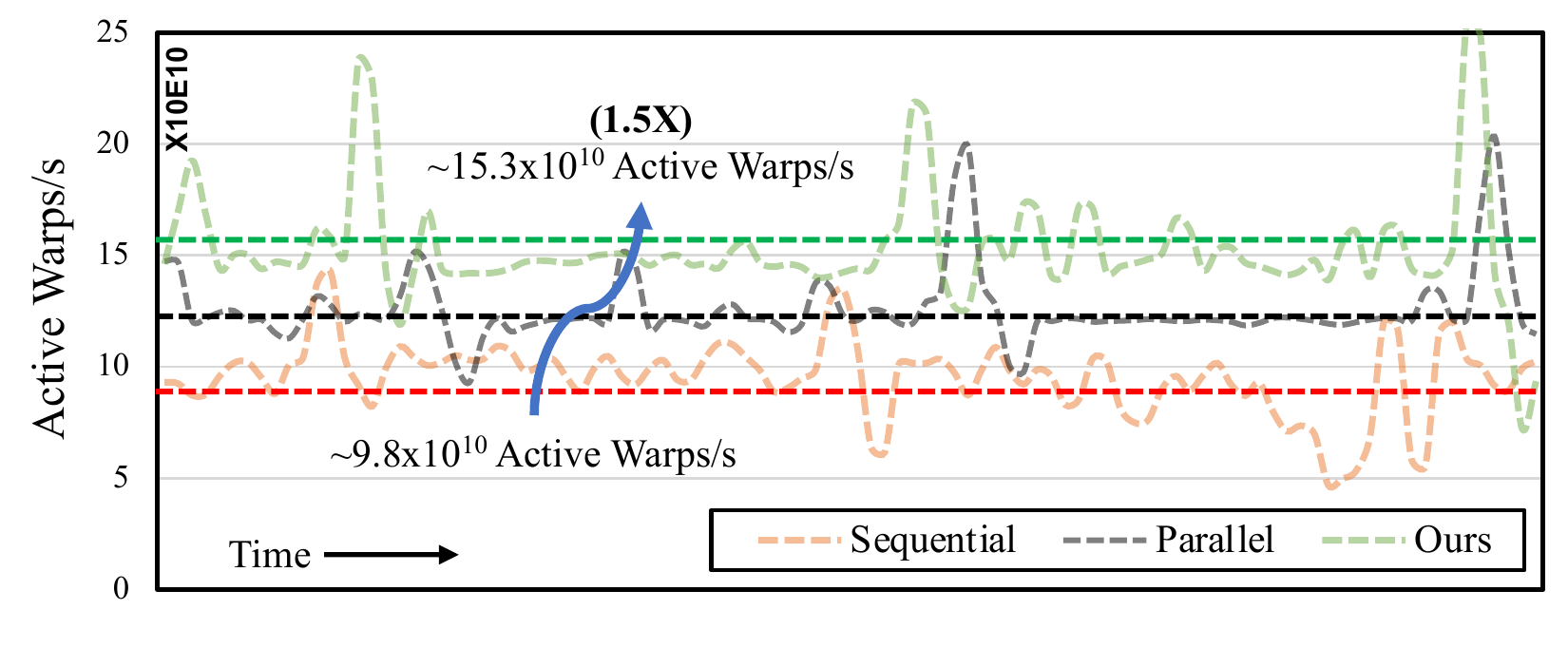}
	\vspace{-5mm}
	\caption{Enhanced GPU Utilization Statistics. The number of active warps per second shows that our schedule could yield continuously better SM utilization.}
	\label{fig:utilization}
\end{figure}

\vspace{1mm}
\noindent \tb{Analysis on GPU Utilization Enhancement}: We further profiled and checked the GPU runtime statistics to analyze and compare the overall GPU utilization with different scheduling strategies.
	Fig.~\ref{fig:utilization} demonstrates the utilization statistics comparison between CuDNN-Sequential, Stream-Parallel, and Our scheduling strategy.
	We use the number of active warps per second as an indicator of GPU utilization information~\cite{ios}.
	As is observed, our scheduling strategy averagely obtains $1.5\times$ utilization enhancement than the sequential schedule, which is consistent with our speed-up performance.

\begin{table*}[!tb]
\centering
\vspace{2mm}
\caption{\textbf{Generality Evaluation (BS=1, 224X224, GPU: P6000 w/ Pascal Arch, Latency: ms})}
\renewcommand\arraystretch{1.5}
\setlength{\tabcolsep}{5.7mm}{
\begin{tabular}{cccccc}
\hline
Models      & CuDNN-Seq & TVM-Seq & Stream-Parallel & Ours-Rand        & Ours-Coor        \\ \hline
Alex + VGG + R18 & 5.754     & 5.523   & 4.694           & 4.225 \textbf{(1.36$\times$)}  & 4.126 \textbf{(1.39$\times$)}  \\ \hline
VGG + R18 + R50  & 9.687     & 8.978   & 8.524           & 7.739 \textbf{(1.25$\times$)}  & 7.425 \textbf{(1.30$\times$)}  \\ \hline
R18 + R34 +R50  & 9.884     & 9.352   & 7.714           & 7.031 \textbf{(1.41$\times$)}  & 6.727 \textbf{(1.47$\times$)}  \\ \hline
R18 + R34 + R101 & 14.278    & 13.256  & 11.833          & 11.08 \textbf{(1.29$\times$)}  & 10.463 \textbf{(1.36$\times$)} \\ \hline
R18 + R50 + R101 & 15.785    & 14.631  & 12.32           & 11.246 \textbf{(1.40$\times$)} & 10.711 \textbf{(1.47$\times$)} \\ \hline
\end{tabular}}
\label{table:general}
\vspace{-4mm}
\end{table*}

\subsection{Scalability and Generality Performance}
In this section, the scalability and generality of our automated scheduling framework are evaluated.

\vspace{1mm}
\noindent \tb{Scalability with Varied Number of Tenants}: 
	We evaluate the scalability of our scheduling framework with varied number of model inference on one single GPU.
	Specifically, we test on three settings: $2\times$ models, $3\times$ models, and $5\times$ models with seven multi-tenant combinations in total.

The overall latency is shown in Table~\ref{table:scale}, which reveals that our framework could scale well with the different number of tenants.
	Our framework could consistently obtain $1.3\times$ to $1.7\times$ acceleration than the sequential baseline across all benchmarks.
	Especially, in the five-model combination setting, we achieve the lowest runtime latency $10.42$ ms, which is $7.5$ ms lower than CuDNN-Seq ($17.96$ ms), and $2.4$ ms lower than Stream-Parallel ($12.85$ ms), demonstrating the huge potential of our framework in accelerating practical applications.

\vspace{1mm}
\noindent \tb{Generality with Different GPUs}: 
We then evaluate the generality of our scheduling framework with different GPU platforms.
	We test five multi-tenant settings on a different GPU: NVIDIA P6000 of Pascal architecture.
	The P6000 GPU is the last version before Titan-V and has slightly lower peak computing performance (12.6 vs. 14.9 TFLOPS).
	As the overall performance in Table~\ref{table:general} shows, our scheduling framework also yield significant performance gain ($1.25\times$ to $1.47\times$ acceleration) on the different GPU platform.

\vspace{1mm}
\noindent \tb{Advantage of Automated Searching}:  The above evaluations demonstrate that our framework could produce an optimal scheduling with better resource utilization and higher runtime speed.
	In addition, the experiments results also reflect one of the most promising advantage of our framework -- \tb{easy-to-scale}.
	With the automated search algorithm design, our framework could automatically find the optimal scheduling strategies for varied number of tenants, distinct multi-model combinations, and different GPU platforms, significantly relieving the scheduling complexity and manual tuning efforts.
\begin{figure}[!b]
	\centering
	\vspace{-5mm}
	\includegraphics[width=3.4in]{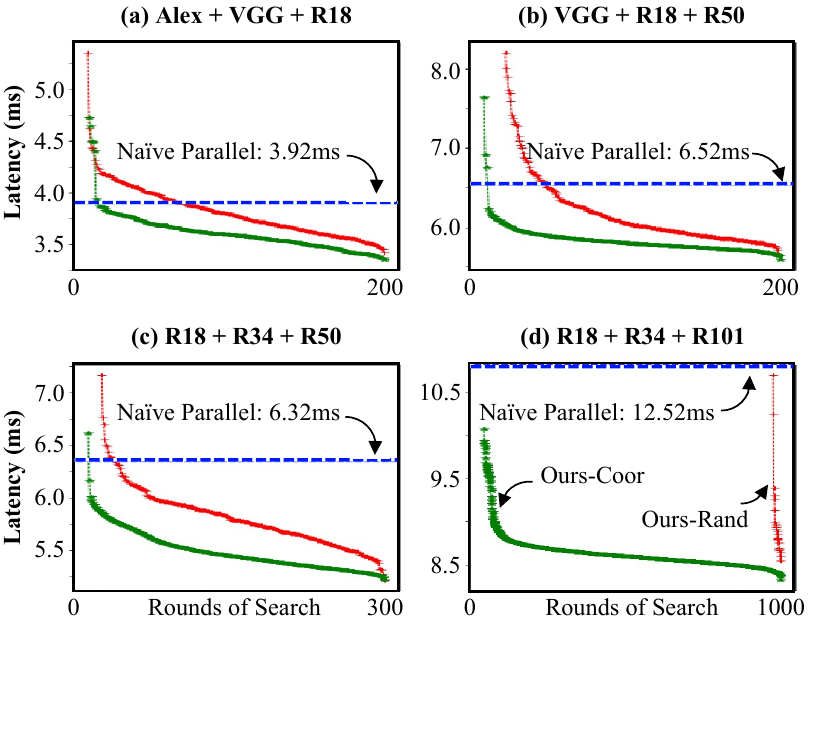}
	\vspace{-15mm}
	\caption{The Search Algorithm Comparison.}\label{fig:search}
\end{figure}

\vspace{-4mm}
\subsection{Search Algorithm Comparison and Overhead Analysis}
In this section, we compare the search algorithms and analyze their introduced off-line running cost.

\vspace{1mm}
\noindent \tb{Search Algorithm Comparison}: 
	Fig.~\ref{fig:search} compares two search algorithms' performance through their searching latency.
	The blue line (\textit{Naive-Parallel}) illustrates the native stream-based scheduling performance.
	The green line (\textit{Ours-Coor}) denotes the scheduling latency with coordinate descent search while the red line (\textit{Ours-Rand}) shows the random search results.
	The same search rounds are conducted in the evaluations.
	The results indicate that the coordinate search generally has better performance than random search in the four multi-tenant conditions.
	Especially, in complex scenarios like Fig.~\ref{fig:search} (d), random search may generate infeasible solutions that are filtered out and leave only few solutions, and thus have slightly worse performance than coordinate descent search.
	Nevertheless, both our search methods outperform the stream-based parallel solution  by a large margin across all cases.

\vspace{2mm}
\noindent \tb{Framework Overhead Analysis}:
Our framework could usually yield near optimal schedule solutions within short search time.
    The framework's running time is demonstrated in Table~\ref{table:overhead}.
    We profile the coordinate descent search with different search rounds from 100 to 1000, which are general settings for most aforementioned multi-tenant scenarios.
    As the results show, our framework's running overhead maintains in the range of ten seconds to several minutes at most.
    Meanwhile, as such automated schedule can be pre-conducted offline given a defined multi-tenant scenario, we consider such offline tuning overhead is highly acceptable.

\begin{table}[!tb]
\centering
\renewcommand\arraystretch{1.5}
\caption{The Framework Running Overhead (Titan-V).}
\vspace{-1mm}
\setlength{\tabcolsep}{2.3mm}{
\begin{tabular}{ccccc}
\hline
\#Search Rounds & 100         & 300         & 500           & 1000          \\ \hline
Alex + VGG + R18    & $\sim$9.8s  & $\sim$28.9s & $\sim$51.4s   & $\sim$1min35s \\ \hline
VGG + R18 + R50     & $\sim$10.3s & $\sim$27.1s & $\sim$48.9s   & $\sim$1min28s \\ \hline
R18 + R50 + R101    & $\sim$16.2s & $\sim$45.3s & $\sim$1min32s & $\sim$2min42s \\ \hline
\end{tabular}}
\label{table:overhead}
\vspace{-3mm}
\end{table}
\vspace{2mm}
\section{Conclusion}
\vspace{2mm}

In this work, we tackle the multi-tenant inference optimization problem on GPU.
    Differently from single-model inference optimization, multi-tenant computation brings significantly higher compute complexity.
To solve such compute complexity, we build an automated scheduling framework for multi-tenant inference optimization.
    Specifically, We first abstract the multi-tenant DNN inference scheduling as  a fine-grained concurrency control problem, and implement the concurrency control by utilizing stream and synchronization based mechanisms.
Based on the problem abstraction, we then formulate the DNN compute graphs and the scheduling factors with a unified IR design.
    Based on that, we formally define the scheduling search space.
In the established scheduling search space, we transform multi-tenant scheduling into an optimization problem and propose an automated ML-based searching algorithm to find the optimal scheduling strategy.
	Experiments demonstrate our method could yield near optimal performance within short time, and meanwhile surpass previous scheduling method by $1.3\times\sim1.7\times$ acceleration.

\newpage
\bibliographystyle{IEEEtran}
\bibliography{ensemble}

\end{document}